\documentstyle[11pt,fleqn]{article}
\date{}
\title{\bf Equations of Motion from Field Equations\\
and a Gauge-invariant Variational Principle \\
for the Motion of Charged Particles}
\author{Dariusz Chru\'sci\'nski\footnotemark\\
        Institute of Physics,
        Nicholas Copernicus University\\
        ul. Grudzi\c{a}dzka 5/7, 87-100 Toru\'n, Poland\\
       and\\
       Jerzy Kijowski\footnotemark\\ Centrum Fizyki Teoretycznej PAN\\
        Aleja Lotnik\'ow 32/46, 02-668 Warsaw, Poland}

\begin{document}

\maketitle

\begin{abstract}

New, gauge-independent, second order Lagrangian for the motion of
classical, char\-ged test particles is proposed. It differs from the
standard, gauge-dependent, first order Lagrangian by boundary terms
only.  A new method of deriving equations of motion from field
equations is developed. When applied to classical electrodynamics,
this method enables us to obtain unambiguously the above, second order
Lagrangian from the general energy-momentum conservation principle.

\end{abstract}

\def\thefootnote{\relax}\footnotetext{$^*$ e-mail: darch@phys.uni.torun.pl}
\def\thefootnote{\relax}\footnotetext{$^\dagger$ e-mail: kijowski@cft.edu.pl}

\section{Introduction}

The motion of classical, charged test particles, in the
classical Maxwell field is derived usually from the {\em
gauge-dependent} Lagrangian function
\begin{equation}
L = L_{particle} + L_{int} = - \sqrt{1-{\bf v}^2} \
( m  - e u^{\mu}A_{\mu}(t,{\bf q}) )\ ,
\end{equation}
where $u^{\mu}$ denotes the (normalized) four-velocity vector
\begin{eqnarray}
(u^\mu) = (u^0 , u^k) :=
\frac 1{\sqrt {1-{\bf v}^2}} (1,v^k)  \ , \label{u-mu}
\end{eqnarray}
and $v^k := {\dot q}^k$ (we use the Heaviside-Lorentz system of units with
the velocity of light $c=1$).

Since the Lorentz force $e u^{\mu} f_{\mu\nu}$ derived from this
Lagrangian is perfectly gauge invariant, it is not clear, why we {\em
have} to use the {\em gauge-dependent} interaction term $e
u^{\mu}A_{\mu}$, with no direct physical interpretation. Moreover, in
this approach equations of motion are not uniquely implied by field
equations. As an example, non-linear forces of the type $u_{\mu}
f^{\mu\lambda} u^\kappa \nabla_\lambda f_{\kappa\nu}$ cannot be {\em a
priori} excluded.

In the present paper we show that the same Lorentz force may be derived
from a gauge-invariant, second order Lagrangian ${\cal L}$:
\begin{equation}
{\cal L} = L_{particle} + {\cal L}_{int} = - \sqrt{1-{\bf v}^2} \
( m  -  a^{\mu} u^{\nu} M_{\mu\nu}^{int}(t,{\bf q},{\bf v}) )\ ,
 \label{calL}
\end{equation}
where $a^\mu := u^\nu \nabla_\nu u^\mu$ is the particle's acceleration.
The skew-symmetric tensor $M_{\mu\nu}^{int}(t,{\bf q},{\bf v})$ is
equal to the amount of the angular-momentum of the field, which is
acquired by our physical system, when the (boosted) Coulomb field ${\bf
f}_{\mu\nu}^{(y,u)}$, accompanying the particle moving with constant
velocity $u$ through the space-time point $y=(t,{\bf q})$, is added to
the background (external) field. More precisely: the total
energy-momentum tensor corresponding to the sum of the background field
$f_{\mu\nu}$ and the above Coulomb field ${\bf f}_{\mu\nu}^{(y,u)}$
decomposes in a natural way into a sum of 1) terms quadratic in
$f_{\mu\nu}$, 2) terms quadratic in ${\bf f}_{\mu\nu}^{(y,u)}$ and 3)
mixed terms. The quantity $M_{\mu\nu}^{int}$ is equal to this part of
the total angular-momentum $M_{\mu\nu}$, which we obtain integrating
only the mixed terms of the energy-momentum tensor. It is proportional
to the particle's electric charge $e$ contained in the particle's
Coulomb field. The interaction Lagrangian ${\cal L}_{int}$ can thus be
expressed in terms of the following quantity
\begin{equation}
Q_\mu:= \frac 1e  u^{\nu} M_{\mu\nu}^{int}  \ ,   \label{Q}
\end{equation}
which is obviously orthogonal to the four-velocity $u^\mu$. Therefore,
it has only three independent components. In the particle's rest frame
we have $Q_0=0$ and the three-vector $e Q_k$ equals to the amount of
the static moment acquired by the system when the particle's own field
is added to the background (external) field. Hence, we have from
(\ref{calL}) and (\ref{Q}):
\begin{equation}
{\cal L}_{int} =  \sqrt{1-{\bf v}^2} \
e a^{\mu} Q_\mu (t,{\bf q},{\bf v}) \ . \label{calL1}
\end{equation}

We prove in Section \ref{proof} that, for a given external field
$f_{\mu\nu}$, the new interaction Lagrangian (\ref{calL1}) differs from
the old one by (gauge-dependent) boundary corrections only. Hence, both
Lagrangians generate the same physical theory (although the new
Lagrangian is of second differential order, its dependence upon the
second derivatives is linear; this implies that the corresponding Euler
-- Lagrange equations of motion are of the second order).  The relation
between $L$ and ${\cal L}$ is, therefore, analogous to the one well
known in General Relativity: the gauge-invariant, second order Hilbert
Lagrangian for Einstein equations may be obtained starting from the
first order, gauge-dependent Lagrangian and supplementing it by an
appropriate boundary term.

At this point, our result can be summarized as follows: a physical
interpretation of the interaction Lagrangian $e u^\mu A_\mu$ has been
found.  Up to boundary terms, it is equal to the
interaction-angular-momentum $a^\mu u^\nu M^{int}_{\mu\nu}$. The
question arises whether such an identity is a pure coincidence or is a
result of a universal law of physics.

In the second part of the paper (Sections 3 -- 5) we try to convince
the reader that the second conjecture is correct. In fact, we propose a
new method of deriving equations of motion from field equations. The
method is based on an analysis of the geometric structure of generators
of the Poincar\'e group, related with any special-relativistic,
lagrangian field theory.  This analysis leads us to a simple theorem,
which we call ``variational principle for an observer'' (see Section
\ref{observer}). Applying this observation to the specific case of
classical electrodynamics, we show how to derive, in principle,
equations of motion from field equations.  This derivation is based on
a following heuristic idea: a point-like particle has to be understood
as an approximation of an extended particle, i.~e.~of an exact, stable,
soliton-like solution of a hypothetical theory of matter fields
interacting with electromagnetic field. To prove that the Maxwell
theory (describing the free field outside of the ``soliton's
strong-field-core'') implies equations of motion for the solitons
themselves, we need several qualitative assumptions of heuristic
nature, about their stability. Under these assumptions the
gauge-invariant Lagrangian describing the motion of point particles is
{\em unambiguously derived} from the general invariance principles of
the theory.

The mathematical status of the above derivation is, therefore, similar
to the Einstein -- Infeld -- Hoffmann derivation of equations of motion
from field equations in General Relativity. It does not depend upon a
particular model which we take for the description of the gravitating
body under consideration (e.~g.~a hydrodynamical or an elastomechanical
model). The derivation is valid for any {\em stable} body and enables
us to describe (in a good approximation) its motion in a
model-independent way, as a geodesic motion of a point-particle.
Hence, even if we have at the moment no realistic mathematical theory
describing the interior of a star and fulfilling all the necessary
assumptions, we can expect that the above equations of motion are valid
for relatively stable objects. The present paper shows that a similar
argumentation is possible also in electrodynamics: Lorentz force acting
on test particles does not need to be postulated as an independent
physical law, but is implied by the geometry of Maxwell field, provided
one accepts the existence of the hypothetical fundamental theory of
matter fields, admitting sufficiently stable soliton-like solutions.

The above result follows immediately from the consistent theory of
interacting particles and fields (cf. \cite{KIJ}, \cite{Kij-Dar}),
called {\em Electrodynamics of Moving Particles}. All the formulae of
the present paper can be derived directly from the above theory in the
{\em test particle limit} (i.e. $m \rightarrow 0$, $e \rightarrow 0$
with the ratio $e/m$ being fixed). The present paper, however, does not
rely on this, much more general, context. The consistent theory of test
particles is constructed independently.

The paper is organized as follows.  Section \ref{proof} contains a
direct proof that our new Lagrangian differs from the standard,
gauge-dependent one by boundary terms only.  Section \ref{observer}
contains our basic geometric observation concerning any relativistic,
Lagrangian field theory, which makes our approach possible. Using it,
we give in Sections \ref{particle} and \ref{renormalization} the
derivation of our variational principle. In particular, the
renormalization procedure defined in Section \ref{renormalization}
depends upon the stability assumptions (which may be relatively
difficult to check for any specific mathematical model). Accepting
these assumptions on a heuristic level we obtain our theory of point
particles uniquely, as a necessary consequence of energy-momentum
conservation.

In a forthcoming paper we are going to present the
gauge-invariant Hamiltonian structure of the above theory.

\section{Equivalence between the two variational principles}
\label{proof}

The easiest way to prove the equivalence consists in rewriting the
field $f_{\mu\nu}$ in terms of the electric and the magnetic induction,
using a special accelerated reference system, adapted to the particle's
trajectory $\zeta$, which we define in the sequel. The system will be
also very useful to formulate our ``variational principle for an
observer'' in the next Section.

We begin with any laboratory coordinate system $(y^\mu) = (y^0,y^k)$ in
Minkowski space-time $M$ and parameterize the trajectory $\zeta$ with
the laboratory time $t=y^0$. Let $y^k = {\bf q}(t)$ be the
corresponding coordinate description of $\zeta$. At each point of the
trajectory we choose an orthonormal tetrad $({\bf e}_\mu)$, such that
its element ${\bf e}_0$ is tangent to $\zeta$, i.~e.~is equal to the
four-velocity vector $u$. Take now the unique boost transformation
relating the laboratory time axis $\partial/\partial y^0$ with the
observer's time axis ${\bf e}_0$.  We define the vector ${\bf e}_k$ by
transforming the corresponding $\partial/\partial y^k$ -- axis of the
laboratory frame by {\em the same} boost.

It is easy to check (cf. \cite{Kij-Dar}) that the above definition
implies the following, explicit formula:
\begin{equation}
{\bf e}_k = \frac {v_k}{\sqrt{1 - {\bf v}^2(t)}}
\frac {\partial}{\partial y^0} +
\left( {\delta}^l_k + \varphi ({\bf v}^2)v^l v_k \right)
\frac {\partial}{\partial y^l} \ , \label{ek}
\end{equation}
where we denote $\varphi (z):= \frac 1{z}
\left( \frac {1}{\sqrt{1 - {z}}} \ - 1 \right) =
\frac 1{\sqrt{1 -z} (1 + \sqrt{1-z} ) } $.

Finally, we parameterize space-time points by four coordinates $(t,x^k)$:
\begin{equation}
y(t,x):= (t,{\bf q}(t)) + x^k {\bf e}_k (t) \ . \label{y-e}
\end{equation}
Using (\ref{ek}) we obtain the following relation
between our curvilinear parameters $(t,x^k)$ and the laboratory
(Lorentzian) coordinates $(y^\mu)$:
\begin{eqnarray}
y^0(t,x^l) & := &
 t +  \frac {1}{\sqrt{1 - {\bf v}^2(t)}} \ x^l v_l(t)
\ ,\nonumber
\\ y^k(t,x^l) & := & q^k(t) +
\left( {\delta}^k_l + \varphi ({\bf v}^2)v^k v_l \right) x^l
 \ . \label{embedding}
\end{eqnarray}

The above formula may be used as a starting point of the entire proof.
To rewrite field equations with respect to this system we need to
calculate the components of the flat Minkowskian metric in our new
coordinates. We see from (\ref{y-e}) that, for a given value $t$,
parameters $(x^k)$ are cartesian coordinates on the 3-dimensional
hyperplane $\Sigma_t$, orthogonal to $\zeta$ at the point $(t, {\bf
q}(t))$. Hence, we get $g_{kl} = \delta_{kl}$ for the space-space
components of the metric.

The remaining information about the metric is carried by the lapse
function and the shift vector, which again may be easily calculated
from formula (\ref{embedding}):
\begin{eqnarray}
N & = & \frac 1{\sqrt{-g^{00}}} = \sqrt{1-{\bf v}^2} \ (1+a_ix^i)\ ,
\label{lapse} \\
N_m &= & g_{0m} =  \sqrt{1-{\bf v}^2} \ \epsilon_{mkl}\omega^k x^l \ .
\label{shift}
\end{eqnarray}
Here, by $a^i$ we denote the rest-frame components of the particle's
acceleration. They are given by formula:
\begin{eqnarray}      \label{e0}
\frac d{d\tau} {\bf e}_0 = \frac d{d\tau} u
= a^i {\bf e}_i \ ,
\end{eqnarray}
where $\tau$ is the proper time along $\zeta$. A straighforward
calculation gives us (formulae (\ref{u-mu}) and (\ref{ek})) the
following value:
\begin{eqnarray}
a^i = \frac {1}{1 - {\bf v}^2}
\left( {\delta}^i_k + \varphi ({\bf v}^2)v^iv_k \right)
{\dot v}^k    \label{ak}
\end{eqnarray}
where ${\dot v}^k$ is the acceleration in the laboratory frame.

Moreover, at each point
of $\zeta$ we define the rotation vector $(\omega^j)$ of the tetrad by
the following formula:
\begin{eqnarray}        \label{ei}
\frac d{d\tau} {\bf e}_i = a_i {\bf e}_0 - \epsilon_{ij}^{\ \ k} \omega^j
{\bf e}_k \ .
\end{eqnarray}
Again, straighforward calculation leads to the following expression:
\begin{eqnarray}
{\omega}_m = \frac {1}{\sqrt {1 - {\bf v}^2}}
\varphi ({\bf v}^2) v^k {\dot v}^l {\epsilon}_{klm}\ .
\label{omegam}
\end{eqnarray}

The transformation (\ref{embedding}) is {\em not} invertible.
Coordinates $(x^k)$ are regular parameters on each $\Sigma_t$. But
hyperplanes $\Sigma_t$ corresponding to different values of $t$ may
intersect. Hence, the same physical event may correspond to different
values of coordinates $(t,x^k)$.

Nevertheless, we may describe the free Maxwell field in terms of our
parameters $(t,x^k)$. In particular, we have:
\begin{equation}
\dot{A}_k - \partial_kA_0 =:
f_{0k} = -ND_k + \epsilon_{mkl} N^m B^l\ ,   \label{dotA}
\end{equation}
where $D^k$ and $B^k$ are the electric and the magnetic field on each
$\Sigma_t$, $N$ and $N^m$ are given by (\ref{lapse}) and (\ref{shift})
(for the description of the Maxwell field with respect to an
accelerated reference system see e.~g.~\cite{MTW}).

Let us multiply (\ref{dotA}) by $\frac {x^k}{r^3}$ and integrate this
scalar product over $\Sigma_t$ with respect to the Lebesgue measure
$dx^3$.  We first integrate over the exterior of the sphere $S(r_0)$.
Observe, that in this region the following identity holds:
\begin{equation}
\frac {x^k}{r^3} \partial_k A_0 = \partial_k
\frac {x^k A_0}{r^3} \ .
\end{equation}
Moreover, we have:
\begin{equation}
\frac {x^k}{r^3} \epsilon_{mkl} N^m B^l =
- \sqrt{1-{\bf v}^2} \ \partial_l \left( B^l \frac {\omega_k x^k}{r}
\right) \ .
\end{equation}
Hence, after integration, we obtain in the limit $r_0 \rightarrow 0$:
\begin{equation}    \label{integrals}
\int_{\Sigma_t}\frac{x^k}{r^3}\dot{A}_k\,d^3x + 4\pi A_0(t,0) =
- \int_{\Sigma_t} \frac{x^k}{r^3}ND_k\,d^3x
\end{equation}
($A_0(t,0)$ is the only surface term which survives in the limit, due
to the standard asymptotic behaviour of the field). Observe that the
constant part of the lapse function (\ref{lapse}) does not produce any
contribution to the right-hand side of the above formula, because the
flux of the field $D_k$ through any sphere $S(r)$ vanishes due to the
Gauss law. Hence, we may replace ``$N$'' by ``$\sqrt{1-{\bf v}^2} \
a_ix^i$'' under the integral and obtain
\begin{eqnarray}
\int_{\Sigma_t} \frac{x^k}{r^3}ND_k\,d^3x &=&
\frac {4 \pi}{e} \sqrt{1-{\bf v}^2}\ a_k \int_{\Sigma_t} x^k {\bf D}^n D_n
\, d^3x \nonumber\\  &=&
- \frac {4 \pi}{e} a^k M^{int}_{k0} =
- \frac {4 \pi}{e} a^\mu M^{int}_{\mu 0}  \ , \label{NDk}
\end{eqnarray}
where
\begin{equation}
{\bf D}^n := \frac{e}{4\pi} \frac{x^n}{r^3} \label{bfD}
\end{equation}
is the Coulomb field on $\Sigma_t$, corresponding to the charge $e$.

The lower index ``0'' in our particular system comes from the vector
$\frac {\partial}{\partial t}$ which is proportional to the particle's
velocity $u^\mu$, where the proportionality coefficient $\sqrt{1-{\bf
v}^2}$ is due to the ratio between the proper time and the laboratory
time on the trajectory. This means that $A_0(t,0)$, calculated in our
particular coordinate system, is equal to $\sqrt{1-{\bf v}^2} \ u^\mu
A_\mu (t,{\bf q}(t))$ in any other coordinate system. The same is
true for $M^{int}_{\mu 0}$. We have, therefore:
\begin{equation}
\frac {e}{4 \pi} \int_{\Sigma_t}\frac{x^k}{r^3}\dot{A}_k\,d^3x +
\sqrt{1-{\bf v}^2} \ e u^\mu A_\mu (t, {\bf q}(t)) =
\sqrt{1-{\bf v}^2} \ a^\mu u^\nu M^{int}_{\mu \nu} \ .
\end{equation}
Integrating this identity over a time
interval $[t_1 , t_2]$ we finally obtain
\begin{equation}
\int_{t_1}^{t_2} {\cal L}_{int} =  \int_{t_1}^{t_2} L_{int} -
\frac{e}{4\pi}  \left(  \int_{\Sigma_{t_2}} \frac{x^k}{r^3}{A}_k\,d^3x
- \int_{\Sigma_{t_1}} \frac{x^k}{r^3}{A}_k\,d^3x
\right) \ ,
\end{equation}
i.~e.~both Lagrangians differ by boundary terms only.

We will see in the sequel that the value of $M^{int}_{\mu \nu}$ does
not depend upon the choice of a particular hypersurface $\Sigma_t$
which we have used for integration. Any other $\Sigma$, which
intersects the trajectory at the same point and is flat at infinity
will give the same result. We conclude that there is always a gauge in
which both ${\cal L}$ and $L$ coincide, e.~g.~the gauge in which the
monopole part of the radial component $A_r: = A_k \frac{x^k}{r}$
vanishes.

\section{Variational principle for an observer }
\label{observer}

Consider any relativistic-invariant, Lagrangian field theory (in this
paper we will consider mainly Maxwell electrodynamics, but the
construction given in the present Section may be applied to any scalar,
spinor, tensor or even more general field theory). Choose any
non-inertial observer, moving along a time-like trajectory $\zeta$. We
want to describe the field evolution with respect to the observer's
rest frame. For this purpose we choose the space-time parameterization
defined in the previous Section.

The field evolution with respect to the above {\em non-inertial
reference frame} is a superposition of the following three
transformations:
\begin{itemize}
\item time-translation in the direction of the local time-axis of the
observer,
\item boost in the direction of the acceleration $a^k$ of the observer,
\item purely spatial O(3)-rotation $\omega^m$.
\end{itemize}
It is, therefore, obvious that the field-theoretical generator of this
evolution is equal to
\begin{eqnarray}
H =  \sqrt{1-{\bf v}^{2}}\left( {\cal E} + a^{k}{R}_{k} -
\omega^{m}S_{m} \right) \ , \label{Hamiltonian}
\end{eqnarray}
where ${\cal E}$ is the rest-frame field energy, ${R}_k$ is the rest-frame
static moment and $S_m$ is the rest-frame angular momentum. The factor
$\sqrt{1-{\bf v}^{2}}$ in front of the generator is necessary, because
the time $t=x^0$, which we used to parameterize the observer's
trajectory, is not the proper time along $\zeta$ but the laboratory
time. For any point $(t,{\bf q}(t))\in \zeta$ the values of the
Poincar\'e generators ${\cal E}, {R}_{k}$ and $S_{m}$ are given as integrals
of appropriate components of the field energy-momentum tensor over any
space-like Cauchy surface $\Sigma$ which intersects $\zeta$ precisely
at $(t,{\bf q}(t))$ (due to Noether's theorem, the integrals are
independent upon the choice of such a surface). These values are,
therefore, equal to the components of the {\em total four-momentum}
$p_\mu$ and the {\em total angular momentum} $M_{\mu\nu}$ of the field,
calculated in the observer's co-moving frame ${\bf e}_\mu$,
i.e.~${\cal E} =
- p_\mu u^\mu$, $R_\mu = - M_{\mu\nu} u^\nu$ and $R_\mu u^\mu = 0$.

Given a field configuration, we are going to use the quantity $H$ as a
second order Lagrangian for the observer's trajectory. For this purpose
let us first choose a ``reference trajectory'' $\zeta_0$. Next, for
each point $(t,{\bf q}(t))\in \zeta_0$ let us calculate the
corresponding ``reference values'' of the generators ${\cal E}(t),
{R}_{k}(t)$, $S_{m}(t)$ and insert them into $H$. Finally, consider the
function (\ref{Hamiltonian}) obtained this way as a Lagrangian
depending upon a generic trajectory $\zeta$ {\em via} its velocity $v$
and acceleration ${\dot v}$, according to (\ref{ak}) and (\ref{omegam}).

\noindent
{\bf Theorem}

\noindent
{\em Euler-Lagrange equations derived from the above Lagrangian are
automatically satisfied by the trajectory $\zeta = \zeta_0$.}

This theorem was derived in \cite{Kij-Dar} in a much more general
framework.  Within this framework, it was an obvious consequence of the
invariance of the theory with respect to the choice of an observer.
More precisely, the function ``$-H$'' was proved to be a {\em Routhian
function} playing the role of the Lagrangian with respect to the
observer's degrees of freedom and the Hamiltonian (with opposite sign)
with respect to the field degrees of freedom. For purposes of the
present paper we do not need, however, this larger context. The Theorem
may be checked by simple inspection: Euler-Lagrange equations derived
from the second order Lagrangian (\ref{Hamiltonian}) are automatically
satisfied as a consequence of the field energy-momentum and
angular-momentum conservation (this direct proof was also given in
\cite{Kij-Dar}).

\section{Adding a test particle to the field }
\label{particle}

From now on we limit ourselves to the case of electrodynamics. This
means that the field energy-momentum and angular-momentum are defined
as appropriate integrals of the components of the Maxwell
energy-momentum tensor
\begin{eqnarray}
T^{\mu}_{\ \nu}=
f^{\mu\lambda}f_{\nu\lambda} - \frac 14
 {\delta}^{\mu}_{\nu} f^{\kappa\lambda}f_{\kappa\lambda}
 \ . \label{Maxwell-tensor}
\end{eqnarray}

Suppose now that to a given background field $f_{\mu\nu}$ we add a test
particle carrying an electric charge $e$. Denote by ${\bf
f}_{\mu\nu}^{(y,u)}$ the (boosted) Coulomb field accompanying the
particle moving with constant four-velocity $u$, which passes through
the space-time point $y$. Being bi-linear in fields, the
energy-momentum tensor $T^{total}$ of the total field
\begin{equation}
f_{\mu\nu}^{total} := f_{\mu\nu} + {\bf f}_{\mu\nu}^{(y,u)}
\end{equation}
may be decomposed into three terms: the energy-momentum tensor of the
background field $T^{field}$, the Coulomb energy-momentum tensor
$T^{particle}$, which is composed of terms quadratic in
${\bf f}_{\mu\nu}^{(y,u)}$ and the ``interaction tensor'' $T^{int}$,
containing mixed terms:
\begin{equation}
T^{total} = T^{field} + T^{particle} + T^{int} \label{decomposition}
\ .
\end{equation}

Unfortunately, $T^{total}$ does not lead any longer to a globally
conserved quantity, as in the case of a relativistic invariant field
theory.  Indeed, the relativistic invariance has been broken by the
choice of the electric current localized on the trajectory. Moreover,
due to the Coulomb field's singularity, integrals which would have been
necessary to obtain the generator (\ref{Hamiltonian}) from
(\ref{decomposition}) are ill defined. Hence, techniques introduced in
Section 3 cannot be used directly.

For reasons which will be fully explained in the next Section, we
replace (\ref{Hamiltonian}) by the following, well defined,
``renormalized'' quantity.  To obtain this quantity we will integrate
first two terms of (\ref{decomposition}) over any $\Sigma$ which passes
through $y=(t,{\bf q}(t))$.  Integration of $T^{field}$ (no
singularity) and $T^{int}$ (an $r^{-2}$ -- singularity) is possible and
gives well defined quantities, which we call respectively $H^{field}$
(``background field generator'') and $H^{int}$ (``interaction
generator'').  Because the left-hand side and the first two terms of
the right-hand side of (\ref{decomposition}) have a vanishing
divergence (outside of the particle's trajectory), we conclude that
also $T^{int}$ has a vanishing divergence.  This implies, that the
above integrals are invariant with respect to changes of $\Sigma$,
provided the intersection point with the trajectory does not change
(see \cite{KIJ} for more details).

Unfortunately, the Coulomb tensor $T^{particle}$ has an $r^{-4}$
singularity at $y$ and cannot be integrated. According to the
renormalization procedure defined in \cite{Kij-Dar} and sketched
briefly in the next Section, we replace its integrals by the
corresponding components of the total four-momentum of the particle:
$p^{particle}_\lambda = m u_\lambda$ and the total angular momentum:
$M_{\mu\nu}^{particle} = 0$. We define, therefore, the renormalized
particle generator as follows:
\begin{equation}
H^{particle}_{ren} =  m \sqrt{1-{\bf v}^{2}} \ .
\end{equation}
Consequently, we define the total (already renormalized) generator as a
sum of three terms
\begin{equation}
H^{total}_{ren} = H^{field} + H^{int} + H^{particle}_{ren} \ ,
\end{equation}
where the first term is quadratic and the second term is linear with
respect to the background field  $f_{\mu\nu}$.

Let us observe that the only non-vanishing term in $H^{int}$ comes from
the static moment term $R$ in (\ref{Hamiltonian}), because the mixed
terms in both the energy ${\cal E}$ and the angular momentum $S$ vanish
when intergated over any $\Sigma$. The easiest way to prove this fact
consists in choosing the hypersurface $\Sigma_t$ which is orthogonal to
the velocity $u$ at $(t,{\bf q}(t))$, i.e.  the rest-frame surface (our
integrals do not depend upon the choice of a hypersurface). On this
surface, the Coulomb field ${\bf f}_{\mu\nu}^{(y,u)}$ is spherically
symmetric and carries, therefore, only the monopole component. On the
other hand, the monopole component of the background field $f_{\mu\nu}$
vanishes as a consequence of the homogeneous Maxwell equations (no
charges!). The mixed term in the energy integral is, therefore, a
product of a monopole-free functions and the pure monopole. Hence, it
vanishes after integration. A similar argument applies to the angular
momentum $S$.

Finally, we have defined
\begin{equation}
H^{total}_{ren} = H^{field} -  \sqrt{1-{\bf v}^{2}} \
a^{\mu} u^{\nu} M_{\mu\nu}^{int}(t,{\bf q},{\bf v})
 + \sqrt{1-{\bf v}^{2}} \ m \ ,
\label{H-total}
\end{equation}
where the interaction term is defined as the following integral
\begin{equation}
M^{int}_{\mu\nu}(y) :=
\int_{\Sigma} \left\{
(x_\mu - y_\mu) T_{\nu\lambda}^{int} (x) -
(x_\nu - y_\nu) T_{\mu\lambda}^{int}  (x) \right\}
d\Sigma^\lambda (x) ,
\end{equation}
and $\Sigma$ is any hypersurface which intersects the trajectory at the
point $y=(t,{\bf q}(t))$.

In particular, using the particle's rest-frame and integrating over the
rest-frame hypersurface $\Sigma_t$ we obtain formula (\ref{NDk}) for
$M^{int}_{k0}$.

As we have already mentioned, the quantity $H^{total}_{ren}$ defined
this way, cannot be directly used in the framework of pure
electrodynamics, as the ``observer's Lagrangian''.  In the next Section
we will show, however, that this quantity provides a good approximation
of the total generator $H$ within a more general framework. In this
framework particles are no longer point-like, but are extended objects,
described by matter fields interacting with electrodynamics. We will
conclude, that within this approximation our ``renormalized generator''
may be used to derive an ``approximative trajectory'' of such extended
objects.

\section{Renormalization. Derivation of equations of motion from field
equations}
\label{renormalization}

According to the approach developed in \cite{KIJ} and \cite{Kij-Dar} we
treat the moving particle as a solution of a hypothetical ``fundamental
theory of matter fields interacting with electromagnetic field''. We
assume that such a theory is a relativistic, Lagrangian (possibly
highly non-linear) field theory.  Moreover, we assume linear Maxwell
theory as a limiting case of the above theory, corresponding to
sufficiently weak electromagnetic fields and vanishing matter fields.

We will suppose that the particle, whose interaction with the
electromagnetic field we are going to analyze, is a global
solution of the above field theory, having following qualitative
properties:
\begin{enumerate}
\item it contains a tiny ``strong field region'', concentrated
in the vicinity of a time-like trajectory $\zeta$, which we may call an
{\it approximate trajectory} of the extended particle,
\item outside of this strong field region the matter fields vanish (or
almost vanish in the sense, that the following approximation remains
valid) and the electromagnetic field is sufficiently weak to be
described by linear Maxwell equations.
\end{enumerate}

To be more precise, we imagine the ``particle at rest'' as a stable,
static, soliton-like solution of our hypothetical ``super theory''. The
solution is characterized by two parameters: its total charge $e$ and
its {\it total} energy $m$. The energy {\em is not} concentrated within
the interior of the particle but contains also the part of the energy
carried by its ``Coulomb tail''. This means that $m$ is an {\it already
renormalized mass}, (or {\it dressed} mass), including the energy of
the field surrounding the particle. Within this framework questions
like ``how big the {\it bare} mass of the particle is and which part of
the mass is provided by the {\it purely electromagnetic} energy?'' are
meaningless.  In the strong field region (i.~e.~inside the particle)
the energy density may be highly non-linear and there is probably no
way to divide it consistently into two such components.

Due to relativistic invariance of the theory, there is a 6
parameter family of the ``uniformly moving particle'' solutions
obtained from our soliton via Poincar\'e transformations.

Now, an arbitrarily moving particle is understood as a ``perturbed
soliton''.  This means that it is again an exact solution of the same
``super theory'', with its strong-field-region concentrated in the
vicinity of a time-like world line $\zeta$, which is no longer a
straight line, as it was for ``uniformly moving particles''. Let us
choose an observer who follows this ``approximate trajectory'' $\zeta$.
We know that he automatically satisfies the Euler-Lagrange
equations derived from the second order Lagrangian (\ref{Hamiltonian}),
where ${\cal E}$, $R^k$ and $S^m$ are the quantities calculated for the
complete non-linear theory.

Suppose now, that the particle may be treated as a {\em test particle}.
This means, that the total field outside of the particle does not
differ considerably from a background field $f$, satisfying homogeneous
Maxwell equations. Using this hypothesis we may approximate, for each
point $(t,{\bf q}(t))$, the exact value of (\ref{Hamiltonian}) by the
value of (\ref{H-total}). Indeed, we may decompose the total
energy-momentum tensor of the complete non-linear field as follows:
\begin{eqnarray}
T^{\mu}_{\ \lambda}=
\left( T^{\mu}_{\ \lambda} - {\bf T}^{\mu}_{\ \lambda} \right)
+ {\bf T}^{\mu}_{\ \lambda}  \ .
\label{dec}
\end{eqnarray}
Here, by ${\bf T}^{\mu}_{\ \lambda}$ we denote the total
energy-momentum of the ``super theory'', corresponding to the
``uniformly moving particle'' solution, which matches on $\Sigma$ the
position and the velocity of our particle. Stability of the soliton
means that the the ``moving particle solution'' does not differ
considerably from the ``uniformly moving particle solution'' inside of
the strong field region (i.~e.~inside the particles). Hence, the
contribution of the first term $(T - {\bf T})$ to the integrals ${\cal
E}$, $R^k$ and $S^m$ may be neglected ``inside the particle'', i.~e.~we
may replace it under integration by the purely Maxwellian quantity
\begin{equation}
(T^{total} - T^{particle}) = T^{field} + T^{int} \ .
\end{equation}
As a result, we obtain the first two terms of (\ref{H-total}).
On the other hand, integrating the last term ${\bf T}$ in (\ref{dec}) we
obtain {\em without any approximation} the corresponding value of the
four-momentum of the particle. This way we reproduce the last term of
(\ref{H-total}).

Replacing (\ref{Hamiltonian}) by its approximate value (\ref{H-total})
and using the Theorem we conclude that the trajectory $\zeta$ has to
fulfill Euler-Lagrange equations derived from (\ref{H-total}). Applying
the Theorem to the linear Maxwell theory we conclude, that the
term $H^{field}$ produces Euler-Lagrange equations which are
automatically fulfilled by $\zeta$. Hence, we may drop out this term,
leaving only the remaining two terms. They finally give us our formula
(\ref{calL}) for the Lagrangian of the test particle (the sign has to
be changed because as a Lagrangian we should have taken ``$-H$''
instead of ``$H$'' -- see remark at the end of Section \ref{observer}).


\end{document}